\documentstyle[prl,aps,tighten,multicol,epsfig]{revtex}
\begin{document}

\title{\bf Self-organized criticality in a model of biological evolution 
with long range interactions} 

\author{Pablo M. Gleiser, Francisco A. Tamarit, and Sergio A. Cannas}
 
\address{Facultad de Matem\'atica, Astronom\'\i a y F\'\i sica,
         Universidad Nacional de C\'ordoba, \\
         Ciudad Universitaria, 5000 C\'ordoba, Argentina \\}

\maketitle

\begin{abstract}
In this work we study the effects of introducing long range 
interactions
in the Bak-Sneppen (BS) model of biological evolution.
We analyze a recently proposed version of the BS model
where the interactions decay  as $ r^
{- \alpha}$; in
this way the first nearest neighbors model is recovered in the limit
$ \alpha \rightarrow \infty$, and the random neighbors version for
$ \alpha = 0$.
We study the space and time correlations and analize  how the 
critical
behavior of the system changes with the parameter $\alpha$.
We also study the sensibility to initial conditions of  the model
using the spreading of damage technique. We found that the system
displays several distinct critical regimes as $\alpha$ is varied from
$\alpha=0$ to $\alpha \rightarrow \infty$

\end{abstract}


\begin{multicols}{2}

\narrowtext

In recent years an increasing numbers of systems that
present Self Organized Criticality \cite{soc1,soc2} have been widely 
investigated.
The general approach of statistical physics, where simple
models try to catch the essencial ingredientes responsable for
a given complex behavior has turned out to be very powerful for the study of this kind of problems.
In particular Bak and 
Sneppen \cite{bak&snep} have introduced a simple model which has 
shown to be able to reproduce evolutionary features
such as punctuacted equilibrium \cite{Gould}. 
Altough this model does not intend to give an accurate description of 
darwinian evolution, it catches into a single and very simple scheme 
(it is based on very simple dynamical rules) several features that 
are expected to be present in evolutionary processes, that is, 
punctuated equilibrium \cite{bak&snep}, Self Organized Criticality 
(SOC) \cite{bak&snep} and weak sensitivity to initial conditions 
(WSIC) \cite{Tamarit,Valleriani} (i.e., chaotic behaviour where the trajectories
 depart with a power law of the time instead of exponentially). In 
this sense, one important question arises about the robustness of 
such properties against modifications (i.e., complexifying) of the 
simple dynamical rules on which the model is based. The original 
model, hereafter referred as the first nearest neighbors (FNN) 
version \cite{bak&snep}, includes only nearest neighbors interactions
 in a one dimensional chain. This model presents SOC \cite{bak&snep} 
and weak sensitivity to initial conditions \cite{Tamarit,Valleriani}.
 On the 
other hand, another version of the model with interactions between 
sites randomly chosen in the lattice (and therefore it can be regarded
 as a mean field version of the FNN), hereafter referred as the random 
negihbors (RN) version \cite{bak&snep2}, does not present SOC \cite{Boer}.
 Moreover, it is not expected (and we shall show in this work that 
it is indeed the case) to present WSIC.

Systems of coevolutionary species are expected to have some distance 
decaying interactions, thus lying somehow between the two previous 
schemes. Although not well defined, the concept of ``distance'' 
between species in these scenarios may be regarded as associated to
 some complex network of relationships including competition for 
resources and predator-prey ones, among many others\cite{Havens}.
 In this sense,
 the environmental modifications produced by the extinction of one 
species may be expected to affect many others not directly 
related to it, where the intensity of such influence 
depends on the above mentioned distance.

Along this line, in this letter we will focuse on the robustness 
of the SOC and sensitivity to initial conditions properties of the
 Bak and Sneppen model against the introduction of long-range 
distance dependent interactions. To this end, we consider
a generalization of the model, recently proposed by 
Cafiero et al \cite{Cafiero2} that takes into account long-range interactions
between species that decay as $r^{-\alpha}$, where $r$ represent 
the distance between species (mesured in lattice units, i.e.,  
$r=1, 2,\ldots$ in a chain) and $\alpha>0$ is a parameter that control
s the effective range of the interactions. The major value of this 
generalization, unlike others introduced in the literature 
\cite{vandewalle1,cafiero}, resides in the fact that 
it allows simply to retrieve the 
two above mentioned models by varying continuously the parameter 
$\alpha$: when $\alpha\rightarrow 0$ we  recover the RN version while
 for $\alpha\rightarrow\infty$ we recover the FNN one.

The model consist of an $N$-site linear lattice with periodic
boundary conditions (i.e., a ring of $N$ sites), where each site 
represents a species.  Each species has associated  
a real variable $b_j$, 
 $0 \le b_j \le 1,$ that measures the relative {\em fitness} barrier. 
Starting from a random barrier distribution, at each successive time  
step we identify the smallest barrier $b_j$, and modify it by  
choosing a new  
random value from a uniform distribution.  This change represents 
a jump of a species across its fitness barrier to a mutated
species. This mutation must also affect other species in the chain,
and to take into account this phenomena one defines a neighborhood
which will also be modified in the same way. In Ref. \cite{bak&snep} 
($\alpha\to \infty$)
the  authors considered 
the case in which this neighborhood consists of
the two nearest 
species of the mutating one while in Ref. \cite{bak&snep2}
($\alpha = 0$) the neighborhood consist of $K-1$ 
species chosen at random among all the species of the chain.
   
In order to generalize these models, we choose the neighborhood
(of $K-1$ species) at random with a probability that decays as  
$r_{ij}^{-\alpha}$, where $r_{ij}$ is  the  distance of a given 
species $j$ to the species with the smallest barrier $i$, and  giving 
them new random values chosen 
from a uniform distribution. In this way,for $K=3$ and
 $\alpha \to \infty$ we recover the two nearest neighbors model, 
while for $\alpha=0$ we reproduce the random neighbors 
mean field version.

To determine whether the system attains a self-organized
critical state, we analyze the following quantities: the barrier 
distribution $P(b)$ in the final steady state, the spatial correlation 
$C(r)$ and the first return time distribution $C(t)$. In order to study the sensitivity to initial 
conditions, we calculate the time evolution of the Hamming distance 
$D(t)$ between two different replicas submitted to the same noise 
(damage spreading method).

\begin{center}
\begin{figure}
\epsfig{file=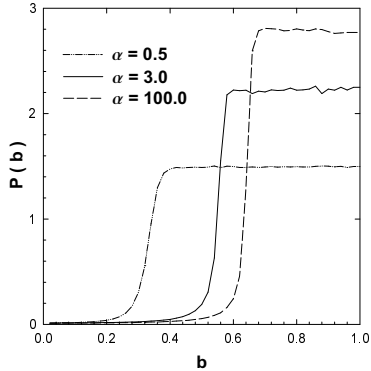,width=50mm}
\caption{Distribution of barrier values for three different values of
$\alpha: 0.5, 3.0$, and $100.0$. All cases have $N=250$.}
\label{fig1}
\end{figure}
\end{center}

\begin{center}
\begin{figure}
\epsfig{file=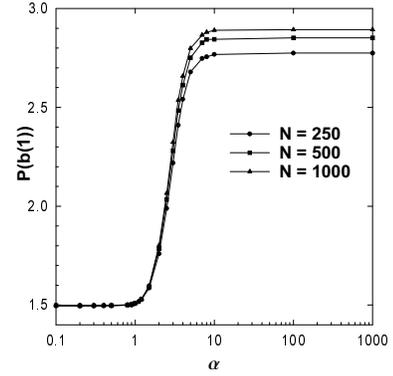,width=50mm}
\caption{$P(b(1))$ vs $\alpha$ for three different system sizes 
$N= 250,500$, and $1000$.}
\label{fig2}
\end{figure}
\end{center}

Since our main interest is to analyze the crossover between 
the limits $\alpha=0$ and  $\alpha \rightarrow \infty$, we will 
restrict ourselves to consider the $K=3$ case.
Latter on we will discuss briefly the effect of increasing $K$.
Figure \ref{fig1} presents the distribution $P(b)$ of barrier values, 
for three different  $\alpha$ values. Note that, independently of 
$\alpha$ the curves are qualitatively similar. 
The typical behavior of these curves can be characterized by the value of 
$P(b(1))$ (i.e, the saturation value of the distribution), which is displayed  in Fig.\ref{fig2}  as a 
function of $\alpha$ for three different system sizes.
We can clearly distinguish three different regimes. 
For $\alpha < 1$ the value of $P(b(1))$ is independent of $N$ 
and the behavior observed for $P(b)$ colapses to the one observed 
in the RN Bak-Sneppen model. 
 There exists some value $\alpha=\alpha_c$ such that for intermediate 
values $1 < \alpha < \alpha_{c}$ the value of  $P(b(1))$ is 
very sensitive to changes in $\alpha$, increasing its value as 
$\alpha$ grows, and finally, for $\alpha>\alpha_c$, $P(b(1))$
 reaches a saturation value,
and we recover the behavior of $P(b)$ for the FNN model when 
$N \rightarrow \infty$. The value of $\alpha_c$ can be roughly 
estimated from numerical extrapolations of the curves 
to $1/N\rightarrow 0$. We obtained that $\alpha_c\approx 4$ for $K=3$ (further analysis of the critical exponents will confirm this
estimation).
\vspace{-0.5cm}

\begin{center}
\begin{figure}
\epsfig{file=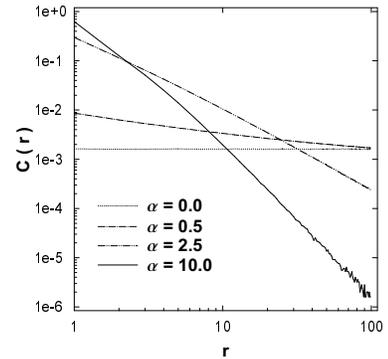,width=50mm}
\caption{Log-log plot of the space correlation $C(r)$, for four
different values of $\alpha:0.0, 0.5, 2,5$, and $10.0$.
 The system size is $N=1000$ }
\label{fig3}
\end{figure}
\end{center}

\begin{center}
\begin{figure}
\epsfig{file=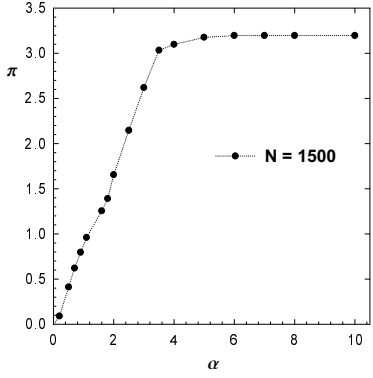,width=50mm}
\caption{Space correlation exponent $\pi$ vs $\alpha$.}
\label{fig4}
\end{figure}
\end{center}

We consider now the spatial and temporal correlations between the 
minimum barriers in order to determine the presence of SOC.
Figure \ref{fig3}  presents in a log-log plot the probability $C(r)$ that the 
minimum barriers  at two succesive updates will be separated by $r$ 
sites. We observed a power law behavior $C(r)\sim r^{-\pi}$ for all 
$\alpha \ne 0$. In Fig. \ref{fig4} we present how $\pi$ changes with 
$\alpha$. When $\alpha=0$ the spatial correlation is constant ($\pi = 0$) as in
 the RN model. As $\alpha$ grows $\pi$ 
increases  until it reaches a saturation value  $\pi = 3.2 \pm 0.2$ 
for $\alpha>\alpha_c$, in agreement with the results observed in the 
FNN model.

\begin{center}
\begin{figure}
\epsfig{file=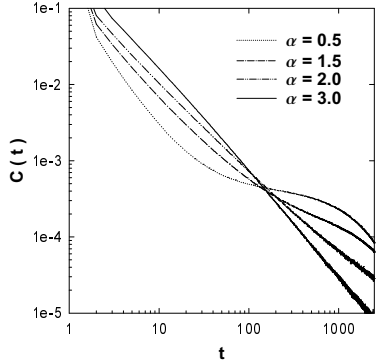,width=50mm}
\caption{Log-log plot of the first return probability $C(t)$,
for four different values of $\alpha : 0.5, 1.5, 2.0$, and $3.0$. 
System size is $N=1000$}
\label{fig5}
\end{figure}
\end{center}

Next we calculate the first return time distribution $C(t)$,  
defined as the  probability that, if a given site is the minimum  at 
time $t_0$, 
it will again be the minimum for the first time at time $t_0 + t$.  
In Fig. \ref{fig5} we present our results for four different values 
of $\alpha$ (0.5, 1.5, 2.0 and 3.0). For $\alpha \geq 2$ 
the first return time clearly presents a power law behavior 
$C(t) \sim t^{-\tau}$, even for finite system sizes. 

\begin{center}
\begin{figure}
\psfig{file=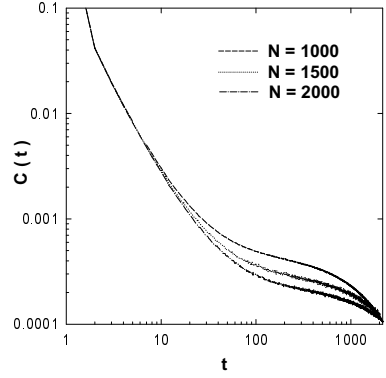,width=50mm}
\caption{Log-log plot of the first return probability $C(t)$, 
for $\alpha = 0.5$,
for three different sizes, $N = 1000, 1500$, and $2000$.}
\label{fig6}
\end{figure}
\end{center}

For $\alpha < 2$ the system displays finite size effects, as  
can be seen in Fig. \ref{fig6} where we present $C(t)$ when 
$\alpha=0.5$ for different system sizes; it is clear that a power law 
decay emerges as $N$ grows.

\begin{center}
\begin{figure}
\epsfig{file=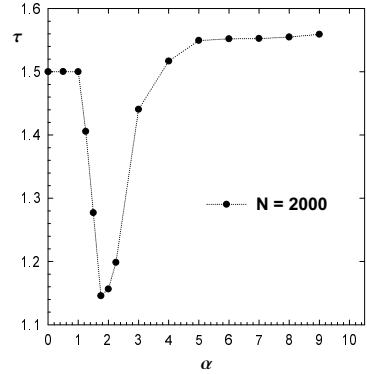,width=50mm}
\caption{First return probability exponent, $\tau$, as a function
of $\alpha$.}
\label{fig7}
\end{figure}
\end{center}

In Fig. \ref{fig7} we show how the first return time exponent 
$\tau$ depends on $\alpha$. Here again we find three  different 
regimes:
For $\alpha < 1$  (unlike the spatial correlation exponent) all
the curves $C_{\alpha}(t)$ colapse and $\tau = 1.5$, displaying the
same behavior found in the RN Bak-Sneppen model with $K=3$, 
where $\tau=3/2$ exactly \cite{Boer}. 
For $1 < \alpha < \alpha_{c}$, the value of $\tau$ strongly 
depends on the value of $\alpha$, with a minimum for $\alpha\approx 1.6$, in agreement with the results of Cafiero et al \cite{Cafiero2}.
For $\alpha > \alpha_{c}$ the value of $\tau$ attains  a saturation 
value $\tau = 1.56 \pm 0.05$ in agreement with 
the value observed in the FNN Bak-Sneppen model where $\tau = 1.6$.

Summarizing the results displayed in these figures, 
for $0 < \alpha < 1$, since $\tau$ presents the same
trivial value observed in the RN Bak-Sneppen model, 
we cannot regard the system as critical \cite{Boer}.
For $1 < \alpha < \alpha_{c}$ the exponents depend 
strongly 
on $\alpha$, and since the exponents are non trivial,
we regard this as a strong indicator of criticality in 
the system. Finally for $\alpha > \alpha_c$ the exponents 
becomes independent of $\alpha$, taking the short range values 
observed in the FNN Bak-Sneppen model.
We have observed that as we increase the number of interacting
sites $K$, the value of $\alpha_c$ decreases, slowly converging 
to $\alpha_c=2$. This behavior reminds that of the one dimensional 
ferromagnetic Ising model with the same type of interactions 
presented here, where the borderline between short and long range
critical regimes is $\alpha=2$ \cite{Dyson}.
 
Next, we study the sensibility to initial conditions of this model
and its dependence on $\alpha$.
To do that, we use 
the spreading of damage technique, which had
previouly been applied to the FNN model \cite{Tamarit,Valleriani}. 
In this particular limit it was shown that the system presents a 
weak sensivity to initial conditions, characterized by a power law
increment, as times goes on, of the Hamming distance between replicas
 of the system. This behavior is reminiscent of those observed at
the edge of chaos in dynamical systems with few degrees of freedom.
The procedure is as follows: given a configuration of $N$ barrier
 values $( \{ b_{j}^{(1)} \} )$ 
in the self-organized critical state, we create a replica of the 
system $( \{ b_{j}^{(2)} \} )$ by choosing 
a site randomly and interchanging the value of this site with the 
value of the site with the smallest barrier.
From then on we use the same random numbers for updating the 
barrier values in both replicas.

We define the Hamming distance between the two replicas as:
\begin{equation}
D(t) \equiv \frac{1}{N} \sum_{j=1}^{N} | \; b_{j}^{(1)}(t) - b_{j}^{(2)}(t) \; |
\end{equation}
If the Hamming distance goes to zero we say that the system is in a 
frozen phase. On the other hand if the Hamming distance remains 
non zero we say that the system is chaotic in analogy with dynamical 
systems.

Regarding the behavior of the average normalized Hamming distance 
$D(N,t)\equiv \left< D(t) \right> / \left< D(1) \right> $, we observed two different 
regimes as $\alpha$ is varied. For $\alpha<2$, $D(N,t)$ reaches a 
saturation value $D(N,\infty)$ in just one step. 
The quotient $D(N,\infty)/N \rightarrow 0$ when $N\rightarrow\infty$, 
and therefore the system does not present sensibility to the initial 
conditions.

\begin{center}
\begin{figure}
\epsfig{file=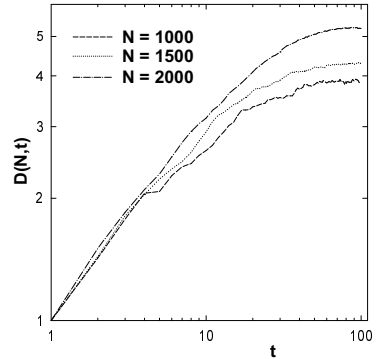,width=50mm}
\caption{Normalized Hamming distance $D(N,t)$ for $\alpha = 2.5$ and
 three system sizes, $N=1000, 1500$, and $2000$.}
\label{fig8}
\end{figure}
\end{center}

The typical temporal behaviour of $D(N,t)$ for $\alpha \geq 2$ is displayed
 in Fig.\ref{fig8} for $\alpha=2.5$ and three different system sizes
(the results presented 
correspond to averages over $500*N$ realizations). We see that 
$D(N,t)\sim t^{\delta}$ when $t\ll N$ and it saturates into a system 
size dependent value for large times, clearly showing WSIC
 \cite{Tamarit,Valleriani}.  
Moreover, we verified  for $\forall \alpha \geq 2$ the finite size scaling 
behavior \cite{Tamarit}:

\begin{equation}
D(N,t) \sim N^{z\delta} \; F\left(\frac{t}{N^z}\right)
\label{dn}
\end{equation}

\noindent where $\delta = 0.40$ \cite{Valleriani2}, and $z=1.7 \pm 0.2$ is the 
dynamical exponent defined by 
$t_{s} \sim N^z$, $t_{s}$ being the value of $t$ at which the 
increasing regime crosses over onto the saturation regime 
\cite{Wang} (given by the
 intersection of the linearly increasing branch of the curve and the 
 horizontal  branch). Both exponents are independent of $\alpha$.

Concluding, we have studied how long range interactions 
affects the criticality of the stationary  
state of our model and its sensibility to the initial conditions. 
Concerning  the SOC, we observed three different regimes depending 
on $\alpha$.
For $\alpha>\alpha_{c}$ we can speak about a short-range critical 
regime, where the system presents SOC. Moreover, we observe 
that this property displays universality, in the sense that most of 
the associated critical exponents are independent of $\alpha$. 
For $0\leq\alpha\leq 1$ the system does not present SOC, although 
$C(r)$ displays a non-trivial power law decay with $r$,
unlike 
the RN model for wich $C(r)$ is constant.  
Moreover, all the relevant state functions or distributions become 
independent of $\alpha$. This behavior has already been observed in 
a variety of systems with long-range interactions, both related to 
equilibrium \cite{Cannas1,Cannas2} and non-equilibrium properties 
\cite{Cannas3,Anteneodo}. In all these systems it has been observed 
that the mean field behavior becomes dominant when 
$0\leq\alpha\leq d$, $d$ being the dimensionality of the underlying 
lattice. Hence, in our case we can speak about a ``mean-field'' 
(non-critical) regime, i.e., that of the RN model. 
Finally, for 
$1\leq\alpha\leq\alpha_c$ we have a long-range critical regime, 
where the system presents non-universal SOC, i.e., the associated critical
exponents depend strongly on $\alpha$.

Concerning the sensibility to the initial conditions we observed 
two regimes: one for $0 \leq \alpha < 2$ where the system does not
present sensibility to the initial conditions of any type, while
for $\alpha >2$ it displays universal WSIC, in the sense that
the exponents of the scaling law (\ref{dn}) are independent of $\alpha$. 

We see that, at least one of the borderline values (and probably all
of them)
 that separate the different regimes  seems to be directly related to 
the dimensionality of the system. Hence, such dimensionality appears 
as a fundamental parameter to determine the robustness of the model 
against variations in the range of the interactions.

Fruitful discussions and suggestions from  S. Boettcher are acknowledged.
This work was partially supported by the following agencies:
CONICOR (C\'ordoba, Argentina), Secretaria de Ciencia y Tecnolog\'{\i}a 
de la Universidad Nacional de C\'ordoba and CONICET (Argentina).


\end{multicols}

\end{document}